\documentclass[aps,prb,twocolumn,groupedaddress,floatfix,fleqn,showpacs]{revtex4}
\usepackage{amssymb}
\usepackage[dvips]{hyperref}
\usepackage[dvips]{graphicx}
\usepackage[T1]{fontenc}

\begin{document}


\title{Microscopic origin of exchange bias in core/shell nanoparticles}


\author{\`Oscar Iglesias}
\email{oscar@ffn.ub.es}
\homepage{http://www.ffn.ub.es/oscar}
\author{Xavier Batlle}
\author{Am\'{\i}lcar Labarta}
\affiliation{Departament de F\'{\i}sica Fonamental, Universitat de Barcelona, Diagonal 647, 08028 Barcelona, Spain}


\date{\today}

\begin{abstract}

We report the results of Monte Carlo simulations with the aim to clarify the microscopic origin of exchange bias in the magnetization hysteresis loops of a model of individual core/shell nanoparticles. Increase of the exchange coupling across the core/shell interface leads to an enhancement of exchange bias and to an increasing asymmetry between the two branches of the loops which is due to different reversal mechanisms. A detailed study of the magnetic order of the interfacial spins shows compelling evidence that the existence of a net magnetization due to uncompensated spins at the shell interface is responsible for both phenomena and allows to quantify the loop shifts directly in terms of microscopic parameters with striking agreement with the macroscopic observed values.
\end{abstract}

\pacs{75.60.-d,05.10.Ln,75.50.Tt,75.60.Jk}

\maketitle


\section{Introduction}
Proximity between a ferromagnetic (FM) and an antiferromagnetic (AFM) material leads to interesting effects that result from the structural modification and competition of different magnetic orderings at the interface between them. 
In particular, the exchange coupling at a FM/AFM interface may induce unidirectional anisotropy in the FM below the Ne\'el temperature of the AFM, causing a shift in the hysteresis loop, a phenomenon known as exchange bias (EB). 
Although the first observations of this phenomenon, dating back five decades ago \cite{Meiklejohn_pr57}, were reported on oxidized nanoparticles, most of the subsequent studies have focused on layered FM/AFM structures \cite{Nogues_jmmm99,Berkowitz_jmmm99} because of their application in advanced magnetic devices \cite{Prinz_Science98}. However, in recent years, the study of EB in nanoparticles and nanostructures \cite{Nogues_physrep05} has gained renewed interest since it has been shown that control of the core/shell interactions or of the exchange coupling between the particle surface and the embedding matrix can increase the superparamagnetic limit for their use as magnetic recording media \cite{Skumryev_nature03}. Several experiments on different nanoparticle systems with oxidized shells have studied the size \cite{Gangopadhyay_jap93,Luna_nano04}, temperature \cite{Peng_prb00,Punnoose_prb01} and cooling field dependence of the EB field \cite{DelBianco_prb04}, as well as training effects \cite{Zheng_prb04}. However, the interpretation of the results may be hindered by collective effects and interactions with the embedding matrix since, up to date, no EB experiment has been conducted on a single particle, which would allow to confront the results with the existing models.
\begin{figure}[bhp]
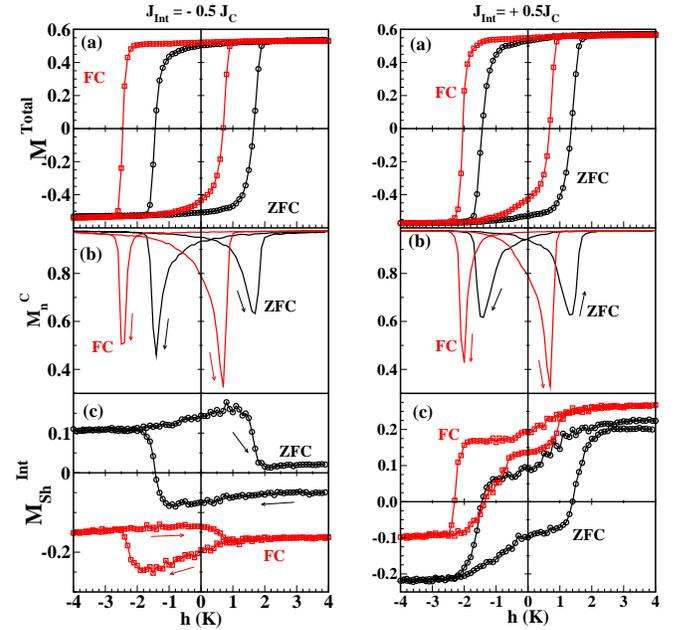

\includegraphics[width=0.49 \columnwidth]{Fig1a.eps}
\includegraphics[width=0.49 \columnwidth]{Fig1b.eps}
\caption{\label{Fig_1}(Color online) Hysteresis loops for a particle with radius $R=12\,a$  obtained from a ZFC state and after FC down to $T= 0.1$ in a field $h_\mathrm{FC}$ for $J_\mathrm{Sh}= -0.5 J_\mathrm{C}$ and $J_\mathrm{Int}=-(+) 0.5 J_\mathrm{C}$ in the left (right) column. Panels (a) display the total normalized magnetization component along the field direction. Panels (b) show the average magnetization projection of the core spins along the field axis. Panels (c) show the normalized contributions of the shell spins at core/shell interface to the total magnetization of the loop.}
\end{figure}

Both nanoparticles and layered systems display common phenomenolgy, although in the later case a wider range of experimenal techniques have been used, which have provided deeper knowledge on the microscopic mechanisms that are at the basis of the EB effect. Recently, spectroscopic techniques have provided insight on the structure and magnetic behavior of the interface spins at a microscopic level, demonstrating the crucial role played by uncompensated interfacial spins on EB \cite{Kappenberger_prl03}. 
%
Thus, knowledge of the magnetic structure at the interface has become a subject of primary interest in understanding EB. At difference from layered systems, the interface of core/shell nanoparticles naturally incorporates roughness and non-compensation of the  magnetization, two of the main ingredients for which different assumptions are adopted by the existing models for EB in films \cite{Malozemoff_prb87,Kiwi_jmmm01}. 
Some microscopic models for bilayers have undertaken calculations of EB fields under certain assumptions \cite{Takano_prl97,Scholten_prb05}, numerical studies based on a mean field approach \cite{Almeida_prb02} or Monte Carlo (MC) simulations \cite{Nowak_prb02,Suess_prb03,Lederman_prb04} making different assumptions about the interface. However, only very recently, some works partially addressing the EB phenomenology in nanostructures have been published \cite{MejiaLopez_prb05,Eftaxias_prb05}.  
%
In this article, we will show, through MC simulations based on a simple model of one core/shell nanoparticle, how some of the EB phenomenology to EB is related to exchange coupling at the core/shell interface. Moreover, the direct inspection of the magnetic configurations along the hysteresis loops will allow us to provide a quantitative understanding of the macroscopic loop shifts in terms of microscopic parameters.  

\section{Model}
The nanoparticles considered have spherical shape with total radius $R= 12 a$ ($a$ is the unit cell size) and are made of a FM core surrounded by an AF shell of constant thickness $R_{\mathrm{Sh}}= 3 a$ with magnetic properties different from the core as well as from the spins at the interface between core and shell spins. Taking $a= 0.3$ nm,  such a particle corresponds to typical real dimensions $R\simeq 4$ nm and $R_{\mathrm{Sh}}\simeq 1$ nm and contains $5575$ spins, of which $45$ \% are on the surface. 

The simulations are based on the following Hamiltonian:
\begin{eqnarray}
\frac{H}{k_B}= -\sum_{\langle  i,j\rangle}J_{ij}\ {\vec S}_i \cdot {\vec S}_j
-\sum_{i}k_i\ (S_i^z)^2 -\sum_{i}\vec h\cdot{\vec S_i}\ ,
\end{eqnarray}
where ${\vec S}_i$ are classical Heisenberg spins of unit magnitude placed at the nodes of a sc lattice. The first term is the nearest-neighbour exchange energy, where the value of the exchange constants $J_{ij}$ depends on the spins belonging to different particle regions. At the core, $J_{ij}$ is FM and will be fixed to $J_{\mathrm{C}}= +10$ K. Spins at the shell have AF coupling as corresponding to oxides; here a reduced value of the coupling at the shell with respect to the core, $J_{\rm{Sh}}= -0.5 J_{\mathrm{C}}$, has been set so that the Ne\'el temperature of the AF $T_{N}$ is lower than the Curie temperature of the FM. Since, in real samples, it is difficult to access microscopic information about the coupling at the interface spins [defined as those on the core (shell) with at least one neighbor on the shell(core)], we have considered different values and signs for the exchange coupling at the interface, $J_{\mathrm{Int}}$.

The second term accounts for the local uniaxial anisotropy along the z-axis. The anisotropy constant at the core is fixed to $k_{\mathrm{C}}= 1$ while the value at the shell, $k_\mathrm{Sh}= 10$, is enhanced with respect to $k_{\mathrm{C}}$ due to the reduced coordination of the shell spins. Finally, the last term describes the Zeeman coupling to an external field $H$ applied along the easy-axis direction, which in reduced units reads $\vec h= \mu \vec H / k_B$.
To simulate the hysteresis loops, we use the MC method with a Metropolis algorithm. As for the spin updates, we use a combination of the trial steps which has proved useful for Heisenberg with finite anisotropies as described elsewhere \cite{IglesiasphaB04}. 
Our protocol to simulate exchange bias mimics the experimental one: we first cool the system from a high temperature $T_0> T_N$ disordered phase in constant steps down to the measurement temperature $T= 0.1$ in the presence of a magnetic field $h_{\rm{FC}}= 4$ K applied along the easy-axis direction. Then, the hysteresis loop is recorded using as starting configuration the one obtained after the field cooling (FC) process. The loops are obtained by cycling the magnetic field from $h= 4$ K to $h= -4$ K in steps $\delta h= -0.1$ K and the different quantites averaged during $200$ MC steps per spin at every field. 
 \begin{figure*}[th]
\includegraphics[width=\columnwidth,angle= -90]{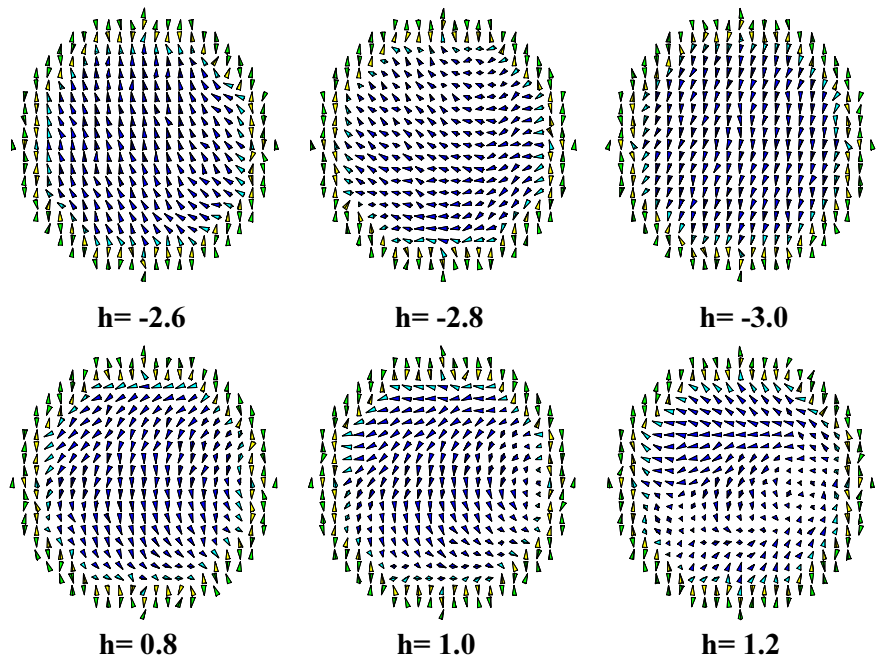}
\caption{\label{Fig_2}(Color online) Snapshots of the spin configurations of a midplane cross section of the particle parallel to the z axis taken at fields of the descending and ascending branches close to the coercive fields ($h_C^-$, $h_C^+$) for the case $J_{\mathrm {Int}}= -0.5 J_\mathrm{C}$ shown in Fig. 1a.
} 
\end{figure*}
\section{Results}
Typical hysteresis loops are shown in Fig. \ref{Fig_1} for two values of the interface coupling $J_\mathrm{Int}/J_\mathrm{C}= -0.5, +0.5$. Compared to the loops obtained from a zero field cooled (ZFC) state, the loops obtained after FC are shifted towards negative field values and have slightly increased coercivity (see Figs. \ref{Fig_1}a), independenly of the sign of the interfacial exchange coupling. In order to gain further insight on the differences between both cases, we have also computed the field dependence of the contribution of interface spins belonging the shell, $M_{\mathrm Sh}^{\mathrm Int}$, to the total magnetization as displayed in the lowest panels of Fig. \ref{Fig_1}. For negative (positive) interface coupling, the interfacial spins at the shell acquire a negative (positive) net magnetization after FC, in both cases higher than the one attained after ZFC, although more pronounced for the negative coupling case. These observations reflect that, after the FC process, a fraction of the interfacial spins ($\approx$15 \% of the interface spins at the shell) have been pinned along a direction compatible with the core/shell exchange interaction, as corroborated also by the vertical shifts in the $M_\mathrm{Sh}^\mathrm{Int}$ loops (to be commented below). This is no longer true for the ZFC case, for which a high fraction of interfacial spins follows the reversal of the FM core as reflected by the change in sign of $M_\mathrm{Sh}^\mathrm{Int}$ along the hysteresis loop.

Interesting enough, we have also noticed an increasing asymmetry of the FC loops with increasing values of the interface coupling, as it is apparent when comparing the descreasing and increasing branches of the loop in the top panels Fig. \ref{Fig_1}a. 
The origin of this asymmetry, is more clearly understood by looking at the average absolute value of the magnetization projection along the field axis through the reversal process, $M_n^{\mathrm C}=\sum_i |\vec{S_i}\cdot\hat{z}|$, as depicted in Fig. \ref{Fig_1}b for the core spins. This quantity presents peaks centered around the coercive fields that indicate deviations of the core magnetization from the applied field direction. 
In the ZFC case, the peaks are centered at similar field values and they are quite narrow and almost symmetric around the minimum. However, for the FC loops, apart for the obvious shift of the peak positions, the decreasing branch peak is symmetric and narrow while the increasing branch one is deeper and asymmetric, enclosing bigger area under the loop curve. 

These observations also indicate that the loop asymmetry reflects different reversal mechanisms in both branches of the hysteresis loops.
This can be corroborated by direct inspection of the spin configurations along the loops, as presented in the main panel of Fig. \ref{Fig_2} for $J_\mathrm{Int}= -0.5J_\mathrm{C}$. As it is evidenced by the sequence of snapshots, the reversal proceeds by quasi uniform rotation along the descending branch, while nucleation of reversed domains at the interface and their subsequent propagation through the core center is basically the reversal process along the ascending branch. Similar asymmetry between the loop branches has been also observed experimentally in bilayers \cite{Fitzsimmons_prl00,Radu_prb03,Eisenmenger_prl05}.
A detailed inspection of the configurations, also reveal the presence of spins at core/shell interface aligned perpendicular to the field direction for intermediate field values (see for example the snapshots for $h= -2.8, 1.0$ in Fig. \ref{Fig_2}). This observation corroborates the interpretation of recent results of small-angle neutron scattering experiments on Fe oxidized nanoparticles in which the anisotropy of the obtained spectra was attributed to the existence of a net magnetic component aligned perpendicularly to the field direction \cite{Loffler_nanostr97,Ijiri_apl05}.

\begin{figure}[tbp]
\includegraphics[width=0.9\columnwidth,angle= 0]{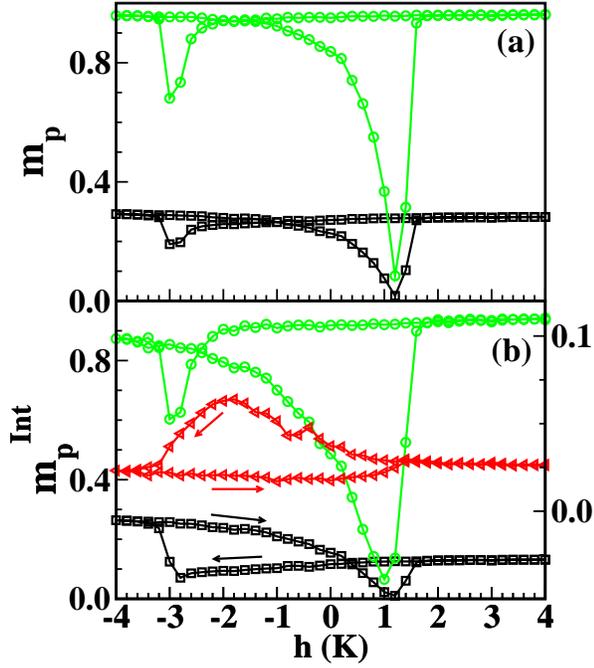}
\caption{\label{Fig_3}(Color online) (a) Shows the field dependence of the average spin projection into the total magnetization direction $m_p$ (squares) and the contribution of the core spins (circles). In (b), the contribution of all the interface spins (squares) has been taken into account, while the contributions of core and shell spins at the interface are shown in circles and triangles, respectively.
} 
\end{figure}

The microscopic origin for the different reversal mechanisms can be further claryfied by looking at the behavior of the interface shell spins along the hysteresis loop (see Fig. \ref{Fig_1}c). While in the descending branch there is a considerable amount of unpinned spins that are able to reverse following the core reversal, in the ascending branch, $M_\mathrm{Sh}^\mathrm{Int}$ remains constant (for $J_\mathrm{Int}<0$), an indication that spins at the shell interface remain pinned, hindering uniform rotation of the core but acting as a seed for the nucleation of reversed domains. 
The changes in the magnetic order at the core/shell interface and the presence of domain walls during reversal can be traced by monitoring the value of the average sum of the projection of the spin direction into the direction of the total magnetization vector along the hysteresis loops as 
\begin{eqnarray}
m_p(h) = \frac{1}{N} \sum_{i=1}^N \vec{S_i}(h)\cdot \vec{M_i}(h)\ .
\end{eqnarray}
This quantity should be close to $1$ if the magnetization reversal proceeds by uniform rotation of the spins, since in this case the spins remain parallel to the global magnetization direction. Deviations of $m_p(h)$ from $1$ indicate the formation of non-uniform structures during the reversal process. 
An example of the field variation of $m_p$ computed for all the spins in the particle and for the interfacial spins is shown in Fig. \ref{Fig_3}, where we have plotted separately the contribution of the core spins.  

During the decreasing field branch of the loop, $m_p$ remains quite close to $1$ for the core spins, except for moderate decrease down to $0.7$ for values of $h$ close to the coercive field at this branch, $h_{\mathrm C}^-$ . The sharpness and symmetry of the peak around $h_{\mathrm C}^-$ confirms that the reversal proceeds by uniform rotation. In constrast, during the increasing field branch, an increasing strong deviation of $m_p$ from $1$ starting from negative field values can be clearly observed, reaching its maximum value also near the coercive field of the increasing field branch, $h_{\mathrm C}^+$, where $m_p\simeq 0$. In this case, the observed peak asymmetry is indicative of the nucleation of the non-uniform domains observed in the snapshots of Fig. \ref{Fig_2}. 
These domains are formed at those points of the core interface with weaker values of the local exchange fields, as indicated by the more pronounced departure from $1$ of $m_p^{\mathrm {Int}}(h)$ (see Fig. \ref{Fig_3}b), than those corresponding to the total magnetization (see Fig. \ref{Fig_3}a). 
The variation of $m_p^{\mathrm {Int}}$ for interface shell spins during the decreasing branch indicates the existence of a fraction of shell spins that reverse dragged by the spins at the core, while constancy of $m_p$ in the ascending branch is indicative of spins pinned during the core reversal.

\begin{figure}[t]
\includegraphics[width=0.9\columnwidth]{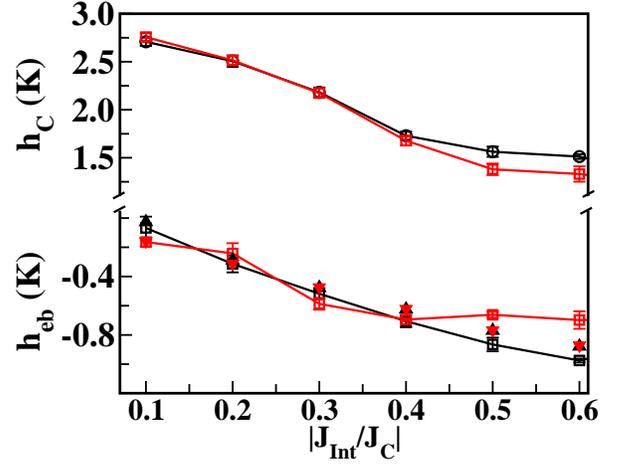}
\caption{\label{Fig_4}(Color online) Variation of the coercive field $h_C$ and the exchange bias field $h_{eb}$ with the exchange coupling constant at the core/shell interface for $J_\mathrm{Int}< 0$ (open circles) and $J_\mathrm{Int}> 0$ (open squares). The exchange bias fields computed from Eq. \ref{Eq_eb} as described in the text are shown with filled symbols for $J_\mathrm{Int}< 0$ (down triangles) and $J_\mathrm{Int}> 0$ (up triangles).
}
\end{figure}

Finally, we have also studied the variation of the coercive field and the EB field [defined as $h_\mathrm{C}= (h_\mathrm{C}^+ - h_\mathrm{C}^-)/2$ and $h_\mathrm{eb}=(h_\mathrm{C}^+ + h_\mathrm{C}^-)/2$ respectively] with the interface exchange coupling $J_\mathrm{Int}$, presented in Fig. \ref{Fig_4} for positive and negative $J_\mathrm{Int}$ values.
For both $J_\mathrm{Int}\gtrless 0$, the values of $h_\mathrm{C}$ and $h_\mathrm{eb}$ are very similar, and a decrease in $h_\mathrm{C}$ and an increase in $h_\mathrm{eb}$ is observed, with a nearly linear dependence for values of $|J_\mathrm{Int}|$ smaller than the exchange coupling at the shell $J_\mathrm{Sh}= -0.5 J_\mathrm{C}$. 
With the increase of $|J_\mathrm{Int}|$, core spins become more coupled to the unpinned shell spins, therefore facilitating the magnetazation reversal with the subsequent decrease in the coercivity. At the same time, increasing $|J_\mathrm{Int}|$ while keeping the values of $J_{\mathrm C}$, $J_{\mathrm Sh}$ and $h_{\mathrm FC}$ constant, results in higher local exchange fields created by the uncompensated spins at the interface, causing an increase of the loop shift.
Let us notice also that the values of the coercive and exchange bias fields obtained are of the correct order of magnitude when expressed in real units. For example, for $J_\mathrm{Int}/J_\mathrm{C}= -0.5$, we obtain $H_\mathrm{C}\approx 0.22$ T and $H_\mathrm{eb}\approx 0.11$ T, which are in agreement with typical values found in studies of oxidized nanoparticles \cite{Skumryev_nature03,Gangopadhyay_jap93,Luna_nano04,Peng_prb00,Punnoose_prb01,DelBianco_prb04,Zheng_prb04,Tracy_prb05}.  

The proportionality of $h_\mathrm{eb}$ to $J_\mathrm{Int}$ should be taken as a hint for the microscopic origin of the loop shifts. As we have mentioned in previous paragraphs, the observed vertical displacements of the loop corresponding to the interface shell spins point to the existence of a net magnetization at the core/shell interface due to uncompensated pinned spins at the shell interface \cite{Nowak_prb02}. If this is the case, the coercive fields after FC can be thought as the sum of the ZFC coercive field $h_\mathrm{C}^0$ and the local field acting on the core spins due to the net interface magnetization of the shell spins, so that they may be computed as 
\begin{eqnarray}
	h_\mathrm{C}^\pm= h_\mathrm{C}^0 + J_\mathrm{Int} M_\mathrm{Int}^{\pm}
\end{eqnarray}
and, therefore, the exchange bias field can be written as
\begin{eqnarray}
	h_\mathrm{eb}= J_\mathrm{Int} (M_\mathrm{Int}^{+}+ M_\mathrm{Int}^{-})/2\ ,
\label{Eq_eb}	
\end{eqnarray}
where $M_\mathrm{Int}^{\pm}=\sum_{i\in \{\mathrm{Int},{\mathrm{Sh}}\}} z_i S_i^z$ is the net magnetization of the interfacial shell spins at the positive (negative) coercive fields $h_\mathrm{C}^\pm$, and $z_i$ is the number of nearest neighbors of spin $i$. 

The last expression establishes a connection between the macroscopic loop shift and microscopic quantities that, although may not be directly measured in an experiment, can be computed independently from the simulation results. The values of $h_\mathrm{eb}$ obtained by inserting the $M_\mathrm{Int}^{\pm}$ values extracted from the Fig. \ref{Fig_1}c in Eq. \ref{Eq_eb} are represented as filled symbols in Fig. \ref{Fig_4}, where we can see that the agreement with the $h_\mathrm{eb}$ values taken from the hysteresis loop shift is excellent within error bars.

\section{Conclusions}
In summary, we have presented simulations of a model for onr core/shell nanoparticle that have revealed an asymmetry in the hysteresis loop due to the different magnetization reversal mechanisms in the two branches. This has been shown to be related to the exchange coupling at the interface $J_\mathrm{Int}$, independently of its sign and of any assumption about the nature of the interface region. The detailed analysis of the changes in the magnetic order of the interfacial spins have also allowed us to demonstrate that macroscopic EB fields can be computed microscopically from the knowledge of interface magnetizations at the coercive fields.

\begin{acknowledgments}
We acknowledge valuable comments on the manuscript by Josep Nogu\'es and also CESCA and CEPBA under coordination of C$^4$ for computer facilities. This work has been supported by the Spanish MEyC through the MAT2003-01124 project and the Generalitat de Catalunya through the 2001SGR00066 CIRIT project.
\end{acknowledgments}

\end{document}